\documentclass[letterpaper]{article}
\usepackage{aaai}
\usepackage{times}
\usepackage{helvet}
\usepackage{courier}
\usepackage{mathtools}
\usepackage{amsfonts}
\usepackage{xcolor}
\usepackage{bm}
\usepackage{subcaption}
\usepackage{graphbox}
\usepackage{balance}
\title{Unsupervised Image Classification by Ideological Affiliation from User-Content Interaction Patterns}
\author{Xinyi Liu, Jinning Li, Dachun Sun, Ruijie Wang, Tarek Abdelzaher\\
Department of Computer Science, University of Illinois at Urbana Champaign\\
201 N. Goodwin Ave., Urbana, IL 61801\\
\AND
Matt Brown, Anthony Barricelli, Matthias Kirchner, Arslan Basharat\\ 
Kitware, Inc.\\
1712 Route 9, Clifton Park, NY 12065\\
}
\begin{document}

\maketitle
\begin{abstract}
\begin{quote}
The proliferation of political memes in modern information campaigns calls for efficient solutions for image classification by ideological affiliation. While significant advances have recently been made on {\em  text\/} classification in modern natural language processing literature, 
understanding the political insinuation in {\em imagery\/} is less developed due to the hard nature of the problem. Unlike text, where meaning arises from juxtaposition of tokens (words) within some common linguistic structures, image semantics emerge from a much less constrained process of fusion of visual concepts. Thus, training a model to infer visual insinuation is possibly a more challenging problem. In this paper, we explore an alternative unsupervised approach that, instead, infers ideological affiliation from {\em image propagation patterns\/} on social media. The approach is shown to improve the F1-score by over 0.15 (nearly 25\%) over previous unsupervised baselines, and then by another 0.05 (around 7\%) in the presence of a small amount of supervision.   
\end{quote}
\end{abstract}

\section{Introduction}
The paper addresses the challenge of {\em unsupervised\/} classification of digital images (e.g., memes) by ideological affiliation. While it is common to describe ideology as a scale from liberal to conservative (or left to right), this is {\em not\/} the definition we adopt in this paper. Ideological divides can arise for many reasons including religious differences, disagreements on historical context, and incompatibilities in the ranking of moral values (e.g., fairness versus loyalty), among many others. The classification algorithm aims to distinguish visual content of two conflicting sides of an ideological divide without a prior understanding of the nature of the underlying divide. The work is motivated by 
the proliferation of memes and other visual aids in marketing~\cite{levinson2001guerrilla}, social movements~\cite{mina2019memes}, and political campaigns~\cite{martinez2016use}, thus generating interest in automating the analysis of ideological and semantic connotations of images~\cite{kiela2020hateful,theisen2021automatic}. Automating machine interpretation of such visual content as memes, however, is arguably a harder challenge than automating scene understanding~\cite{naseer2018indoor,grant2017crowd,xue2018survey} or text understanding~\cite{brown2020language} due to the much less structured nature of the underlying creative process behind meme generation. New memes often utilize unique and subtle juxtapositions of concepts, whose novel nature challenges self-supervised models trained on existing patterns.
This work skirts the problem by exploring an {\em alternative\/} unsupervised approach to  image interpretation. Namely, we use visual similarity and interaction patterns between users and (families of) images to classify images by ideological leaning. 

The idea of exploiting user-content interactions for unsupervised content classification has previously been explored by the authors in the context of classifying text posts~\cite{al2017unveiling,li2022unsupervised,yang2020hierarchical}. It stems from the observation that users interact with content that matches their beliefs. Thus, in an ideological clash, characteristic of today's growing polarization~\cite{lelkes2016mass,petri2021weathering}, different groups of users interact with different content items depending on their ideological leaning. While the classification algorithm need not know the ideological leaning of users ahead of time, it can cluster content by the way it propagates, thereby separating content into ideologically aligned categories.

This paper explores the logical follow-up question: can the same approach be successfully applied to images? How well will it work, and what design parameters are relevant to improving its performance? The underlying reason why a classification approach that relies on observing user-content interactions is of interest lies in that social media (the main digital media where memes propagate in the first place) associate posts with sources. This association allows one to assess ideological similarity among posts by assessing similarity in the sets of users (i.e., sources) who propagate them. This similarity measure is entirely independent of user identity, which can in fact be kept anonymous to the classifier. The unsupervised algorithm, therefore, does not exploit user features (and passed the IRB approval process).\footnote{This work passed the ethical approval process by the institutional review board (the IRB) and was deemed exempt.}  

More specifically, in this paper, we use a modified variational graph auto-encoder, originally proposed for improving feature disentanglement in the latent space~\cite{shao2020controlvae} and subsequently adapted for (ideological) belief representation learning~\cite{li2022unsupervised}. The adapted version, called InfoVGAE, is applied to a bipartite graph of sources and {\em visual assertions\/}, each representing a group of very similar images or memes. The graph is mapped into a latent space where visual assertions of a similar ideological leaning are clustered together. We evaluate the approach based on  image dataset we collected from online controversies over the Russia-Ukraine war. The evaluation shows that the approach is successful at separating two clusters of ideologically distinct images; one represents pro-Kremlin imagery and the other pro-Ukraine imagery. 

\begin{figure*}
    \centering%
    \includegraphics[width=\textwidth]{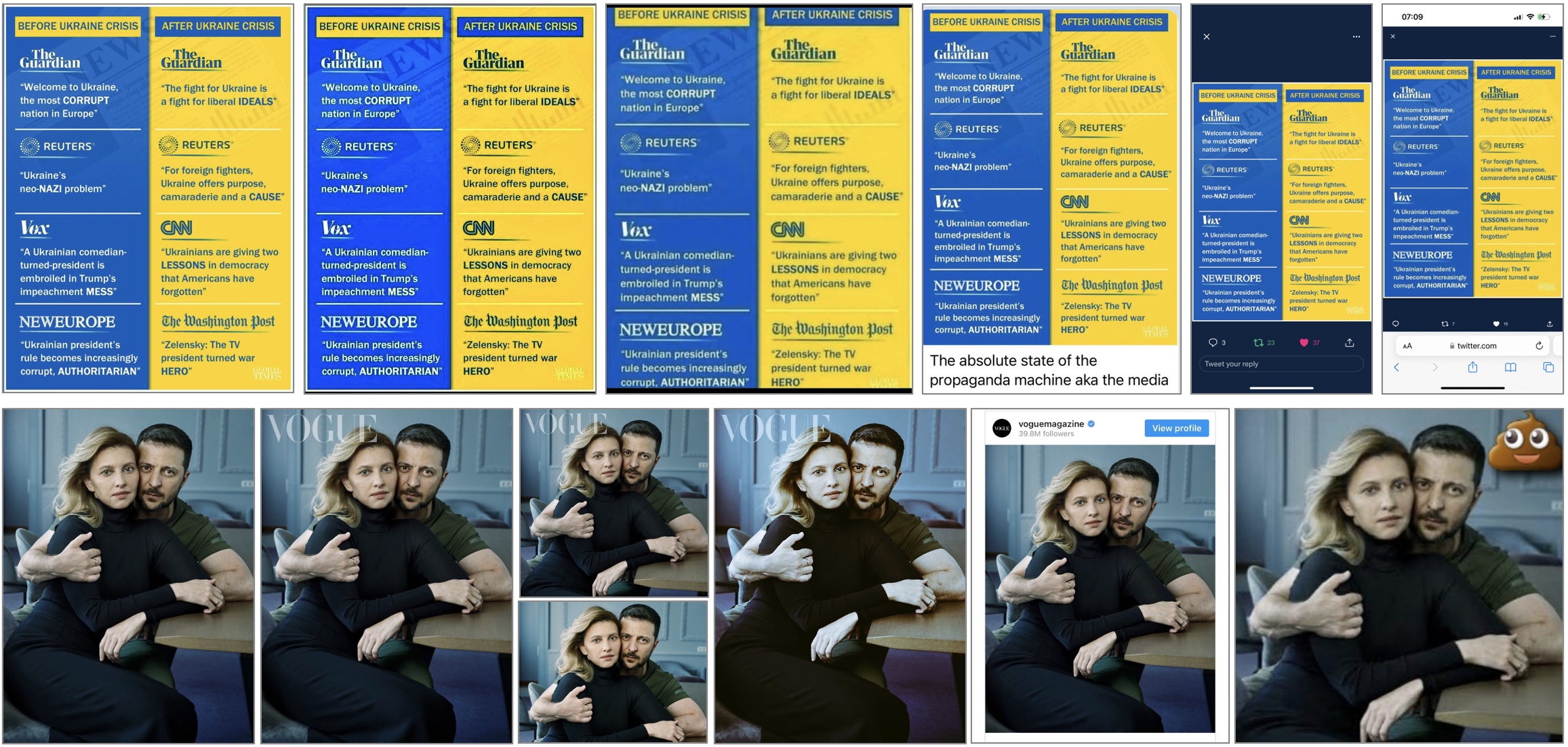}
    \caption{Representative examples of two identified near-duplicate visual assertions in the Russia-Ukraine image dataset.}
    \label{fig:near-duplicate}
\end{figure*}

The rest of the paper is organized as follows. Section~\ref{sec:embedding} reviews some background and presents our InfoVGAE-based image embedding algorithm. Section~\ref{sec:evaluation} describes the data set used for evaluation and presents evaluation results. Section~\ref{sec:discussion} discusses limitations of the current approach and proposes avenues for future work. Section~\ref{sec:related} summarizes related work. The paper concludes with Section~\ref{sec:conclusions}.

\section{Image Embedding and Classification}
\label{sec:embedding}
This section takes the reader through the step-by-step process of (i) identifying visual themes in messages, called {\em visual assertions\/}, (ii) constructing the user-assertion interaction graph from social media data, and (iii) performing self-supervised embedding on the resulting graph into a lower-dimensional ideological space. Semi-supervised extensions are also described. 

\subsection{Identifying Visual Assertions}
\label{sec:visual_model}
In order to identify meaningful user-content interactions, we first need an approach for recognizing (and grouping together) content items that have very similar semantics. Each group of nearly identical items represents essentially the same intended message. Identifying such groups makes it easier to learn how a community propagates similar items. 

We call each semantically-similar group of items an {\em assertion\/}. In this case, we are referring specifically to images. Thus, for the purposes of this work, we define a \emph{visual assertion} as a set of images that share a high degree of similarity. We further assume that a given image is associated with at most one visual assertion, although extensions to many-to-many mappings are possible. Different definitions of image similarity will lead to different interpretations of what a visual assertion represents. In this work, we focus on near-duplicate images (i.\,e., sets of images that were very likely derived from each other through operations like resizing, cropping, recompression, color adjustment, or adding text-overlays, amongst many others). Examples of two visual assertions are shown in Figure~\ref{fig:near-duplicate}. Specifically, we use a keypoint-based approach to identify near-duplicate images, where we declare a pair of images as near-duplicates if there is a sufficient number of matching ORB \cite{rublee2011ORB} keypoints across both images, and if the affine image-to-image mapping estimated with RANSAC \cite{fischler1981ransac} from the set of detected keypoints is empirically sensible. We can then compile the set of visual assertions from the cluster graph obtained from all detected pairs of near-duplicates in the set of candidate images. We narrow down the set of candidate images prior to the near-duplicate detection by excluding pairs of images with a low cosine similarity in the CLIP \cite{radford2021CLIP} embedding space.\footnote{CLIP itself is not suitable for near-duplicate detection as it is generally more broadly indicative of semantic similarity.}

\subsection{Constructing the User-Image Interaction Graph}
\label{sec:graph}
Given an algorithm for clustering similar images into visual assertions, described above, we construct a user-assertion interaction graph as follows:

\begin{itemize}
    \item {\bf Step 1:\/}  Extract the user(s) who posted each individual image in the collected image dataset. This is typically a straightforward look-up of object metadata using the respective social network API.
    \item {\bf Step 2:\/}  Cluster the images based on the method proposed in Section~\ref{sec:visual_model}. Represent each image cluster by a visual assertion node.
    \item {\bf Step 3:\/}  Represent each user by a user node. Link each user node to all visual assertion nodes to which the user contributed images. 
\end{itemize}

\begin{figure}[!ht]
  \centering
  \includegraphics[width=0.49\textwidth]{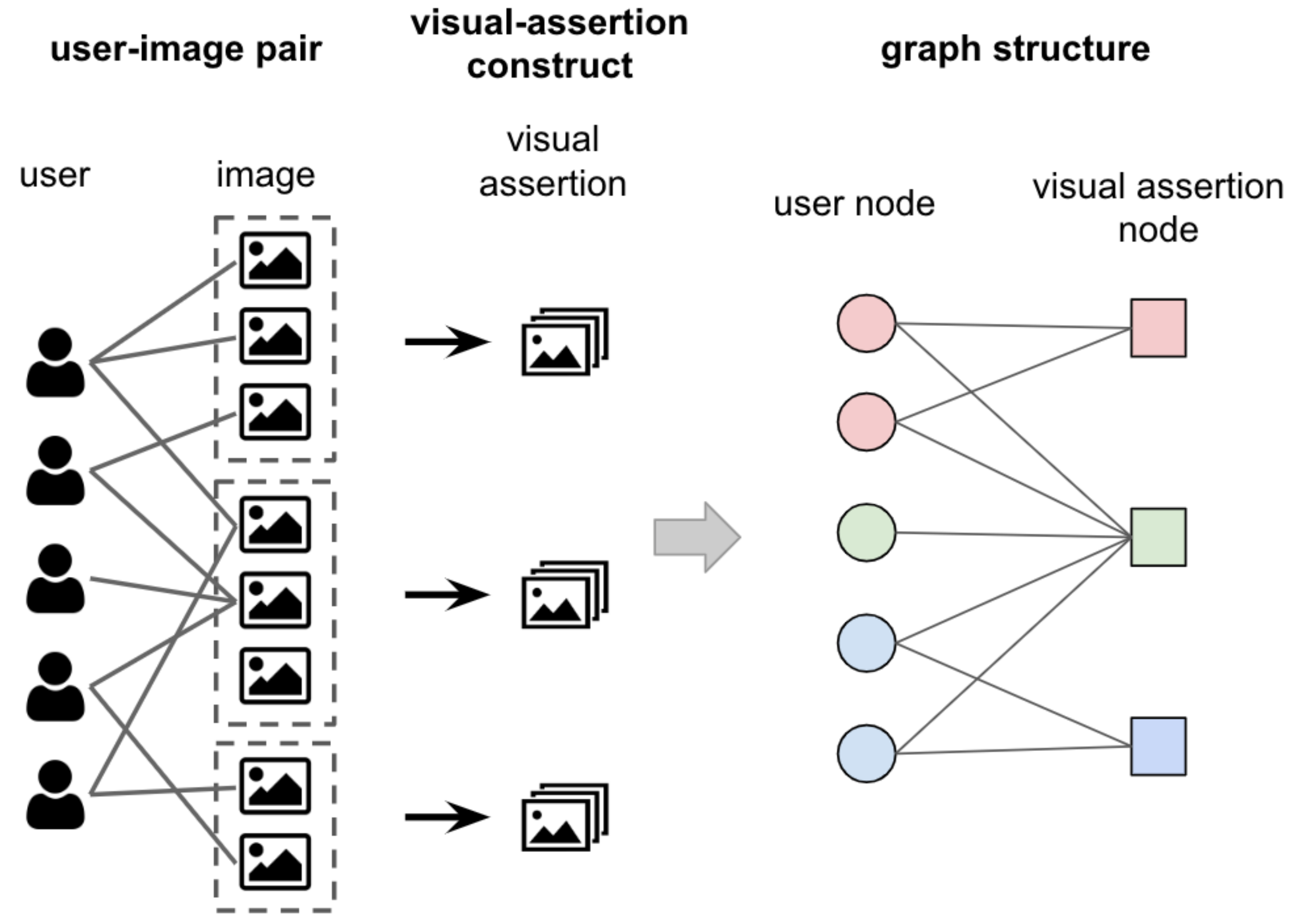}
  \vspace{-5pt}
  \caption{Three steps of converting user-image pairs to a Bipartite Heterogeneous Information Network (BHIN) user-image interaction graph.}
  \label{fig::graph_construct}
  \vspace{-5pt}
\end{figure}

After applying the above procedure, we can model the users and visual assertions by a \textit{Bipartite Heterogeneous Information Network (BHIN)}~\cite{sun2012mining} given by a graph $\mathcal{G}=\{\mathcal{V}, \mathcal{E}\}$, where the number of vertices, $|\mathcal{V}|=N$, is the sum of user and assertion vertices, and the number of edges,  $|\mathcal{E}|=M$, is the number of user-assertion links, as is shown in Figure~\ref{fig::graph_construct}. There are two vertex types in the graph, users and visual assertions. In general, the number of edge types could be
$R$, representing different operation types in the social network, such as posting, commenting, replying, etc. In our implementation, we use only one edge type that represents posting/reposting.

\subsection{Unsupervised Embedding}
\label{sec:embed}
We embed the user-assertion interaction graph, described above, into a lower-dimensional latent space using a version of \textit{variational graph auto-encoders}, called InfoVGAE~\cite{li2022unsupervised}. Each dimension in the latent space represents a different ideological leaning. Thus, in the case of a two-sided conflict, we embed the user-assertion interaction graph into a two-dimensional space. The loss function of the embedding algorithm (i.e., the criterion optimized by the placement of nodes in the latent space) encourages (i) placing pairs of nodes connected by an edge onto the same latent axis, and (ii) placing pairs of nodes with no common edge onto geometrically orthogonal axes. Thus, content propagated by largely different sets of users ends up mapped to different axes, offering the basis for separating different ideological leanings. Below, we review the basic mathematical background on vanilla VGAEs then describe the used~\mbox{InfoVGAE}.

\subsubsection{VGAE Preliminaries:}
A variational graph auto-encoder (VGAE)~\cite{kipf2016variational} is an unsupervised algorithm designed to embed graph-structured data into a lower-dimensional latent space. 
Our VGAE accepts as input the undirected, unweighted graph, $\mathcal{G}=\{\mathcal{V}, \mathcal{E}\}$, constructed in Section~\ref{sec:graph}. The edges of $\mathcal{G}$ are represented by an adjacency matrix (with self-loops), denoted by $\bm{A}$. The nodes are represented by matrix $\bm{X}\in \mathbb{R}^{N\times F}$, called the input feature matrix, where $F$ is the length of each node's feature vector. The typical VGAE consists of two parts, namely, the inference model (Encoder) and the generative model (Decoder), described below. 

\medskip
{\bf A. Inference Model (Encoder):\/}  The encoder exploits an $L$-layer \textit{graph convolutional network} (GCN) as the neural network architecture. The hidden state of nodes is denoted by $\bm{E}^{(l)}\in \mathbb{R}^{N\times d_l}$, where $d_l$ is the dimension of hidden state in the $l^{th}$ layer. $\bm{E}^{(0)} = \bm{X}$ is the input feature. The per-layer computation within the GCN can be represented as $\bm{E}^{(l)}=GCN^{(l)}(\tilde{\bm{A}}, \bm{E}^{(l-1)})$, which receives the hidden state of the previous layer $\bm{E}^{(l-1)}$ and the normalized adjacency matrix $\tilde{\bm{A}}$ as input. More specifically, the GCN layer is formulated~as
\begin{equation}
    \bm{E}^{(l)} = \gamma\left(
    \widetilde{\bm{A}}\bm{E}^{(l-1)}\bm{W}^{(l-1)}
    \right), ~ (2\leq l \leq L-1)
\end{equation}
To formulate the output of the encoder as variational latent space, the output of the last layer of the GCN represents its parameters as the mean and standard deviation vectors $\bm{\mu}_i$ and $\bm{\sigma}_i$. We have $\bm{\mu}=\widetilde{\bm{A}}\bm{G}^{(L-1)}\bm{W}_{\bm{\mu}}^{(L-1)}$ and $\log \bm{\sigma}=\widetilde{\bm{\bm{A}}}\bm{G}^{(L-1)}\bm{W}_{\bm{\sigma}}^{(L-1)}$.
The latent representation matrix $\bm{Z}\in \mathbb{R}^{N\times T}$ is then derived with the re-parameterization trick~\cite{kingma2013auto}, with $T$ as the dimension of the target latent space. We denote $\bm{z}_i$ as the latent vector of the $i^{th}$ node. The inference model is defined as:
\begin{equation}
    q(\bm{Z}|\bm{A}, \bm{X}) = \prod_{i=1}^{N} q(\bm{z}_i|\bm{A}, \bm{X}),
    \label{equ::encoder}
\end{equation}
where $q(\bm{z}_i|\bm{A}, \bm{X}) \sim \mathcal{N}(\bm{z}_i|\bm{\mu}_i, \bm{\sigma}_i^2)$ with $\mathcal{N}$ as the Gaussian Distribution.


\medskip
{\bf B. Generative Model (Decoder):\/} The generative model is given by the inner product between variables. It is formulated as:
\begin{equation}
    p(\bm{A} | \bm{Z}) = \prod_{i=1}^N \prod_{j=1}^{N}
    p(\bm{A}_{i,j} | \bm{z}_i, \bm{z}_j),
\end{equation}
where $p(\bm{A}_{i,j} | \bm{z}_i, \bm{z}_j) = \sigma(\bm{z}_i^T\bm{z}_j)$. 
$\bm{A_{i,j}}$ are the elements of $\bm{A}$ and $\sigma(\cdot)$ is the logistic sigmoid function. Finally, the VGAE model is trained to maximize the variational lower bound,
\begin{equation}
\mathbb{E}_{\bm{Z}\sim q(\bm{Z}|\bm{A}, \bm{X})} \left[\log p(\bm{A} | \bm{Z})\right] - D_{KL}\left[q(\bm{Z}|\bm{A}, \bm{X}) \| p(\bm{Z})\right]
\end{equation}

\subsubsection{InfoVGAE:}
While the vanilla VGAE model, described above, can be applied to learn the node representations from a general graph, we find it helpful to provide certain additional constraints on the learned representation to better separate content by propagation patterns. Specifically, we use a VGAE variant, called the Information-Theoretic Variational Graph Auto-Encoder (InfoVGAE)~\cite{li2022unsupervised}. An InfoVGAE creates a non-negative latent space by replacing the standard Gaussian Distribution with a rectified Gaussian Distribution~\cite{socci1998rectified}. Thus, it models the posterior probability in Equation~\ref{equ::encoder} as $q(\bm{z}_i|\bm{A}, \bm{X}) \sim \mathcal{N}_{+}(\bm{z}_i|\bm{\mu}_i, \bm{\sigma}_i^2)$ where $\mathcal{N}_{+}$ is the rectified Gaussian distribution. This small modification has the profound effect that the only geometrically orthogonal latent representation becomes one where node embeddings are directly aligned with the axes. Thus, a loss function that favors orthogonality among non-interacting sets of items forces them to lie approximately on the axes, leading to an interpretable representation, where each axis maps differently-propagating content (and, hence, a different ideological leaning).  

The InfoVGAE introduces two additional modifications to further improve the embedding by decreasing disentanglement between different dimensions in the latent space (so that each dimension can represent a different ideology).

First, it penalizes the correlation between different embedding axes, by jointly training a discriminator $\Phi$ to minimize the total correlation loss:
\begin{equation}
    \mathcal{L}_{r}(\bm{Z}) = \mathbb{E}_{\bm{Z}\sim q(\bm{Z})}[log(\Phi(\bm{Z})) - log(1-\Phi(\bm{Z}))].
\end{equation}
Second, InfoVGAE explicitly  controls the KL divergence with a PI-controller that manipulates a new parameter $\beta(t)$~\cite{shao2020controlvae}. The final objective of the InfoVGAE is formulated as one of maximizing:
\begin{align}
\begin{aligned}
&\mathbb{E}_{\bm{Z}\sim q(\bm{Z}|\bm{A}, \bm{X})} \left[\log p(\bm{A} | \bm{Z})\right] \\
&~~~~~~~~ - \beta(t) D_{KL}\left[q(\bm{Z}|\bm{A}, \bm{X}) \| p(\bm{Z})\right] - \lambda \mathcal{L}_{r}(\bm{Z})
\end{aligned}
\end{align}
As we show in the evaluation section, the produced embedding directly separates the visual content into (two) clusters aligned with the different axes in the latent space. Each cluster corresponds to a different ideological leaning. 

\begin{figure}[!ht]
  \centering
  \includegraphics[width=0.49\textwidth]{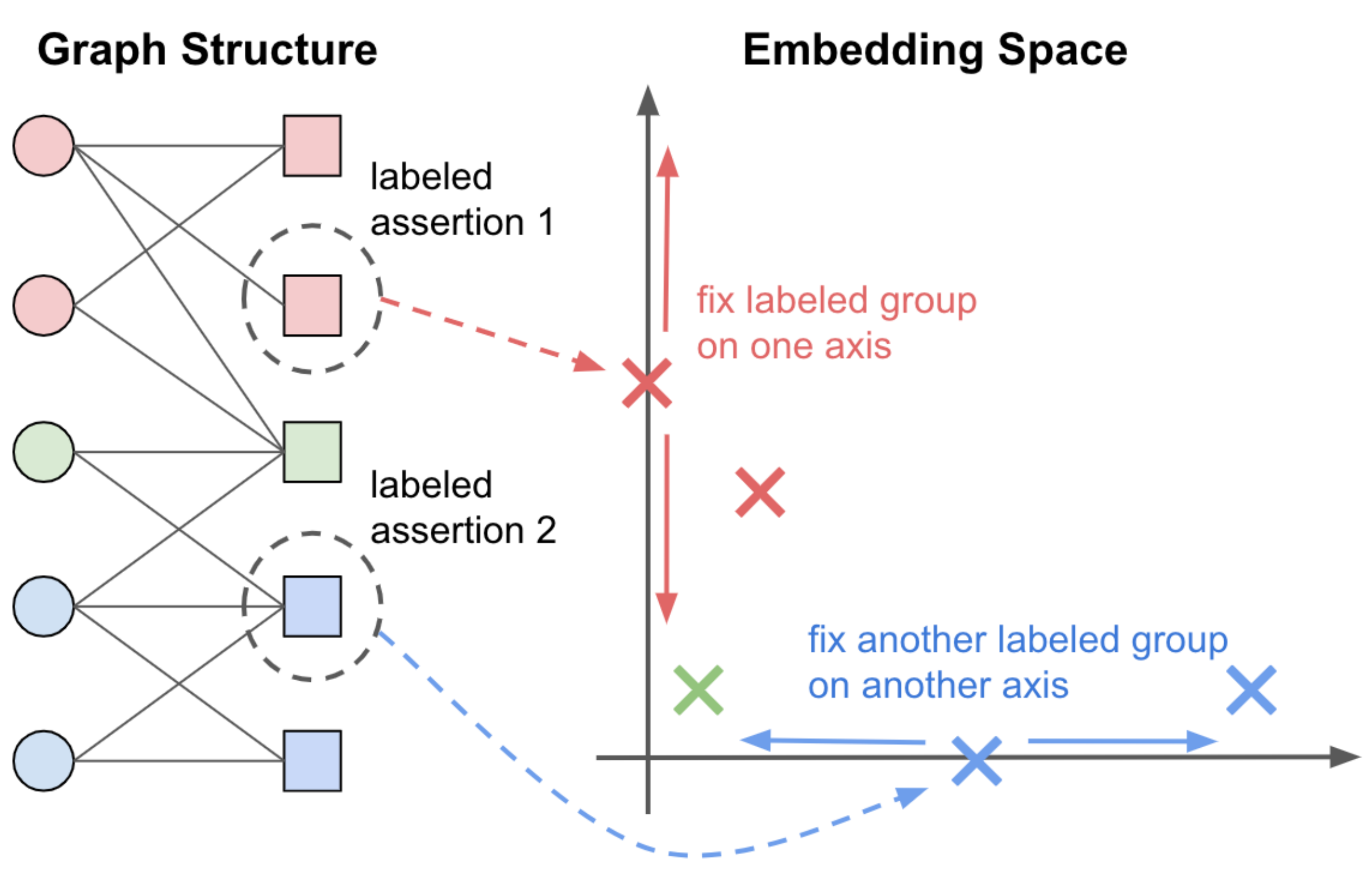}
  \vspace{-5pt}
  \caption{The left part of the figure is the constructed user-image BHIN, where several visual assertion nodes in each ideology group are human-labeled. The embedding of the labeled visual assertions in different ideology groups are fixed on different axis as is shown on the right.}
  \label{fig::semi-supervised}
  \vspace{-5pt}
\end{figure}

While the InfoVGAE is fully self-supervised, in the evaluation section, we additionally explore accuracy gains attained when the it is used in a semi-supervised manner, to leverage situations where some assertions are possible to label (by ideological leaning) ahead of time.
In this scenario, we manually label some of the least popular visual assertions upfront. In the InfoVGAE training stage, since each latent axis corresponds to a different leaning, we augment the loss function with a regularization term that penalizes non-zero values of coordinates of the opposite (i.e., wrong) leaning for each labeled assertion, essentially forcing it to lie on a specific axis, as is shown in Figure~\ref{fig::semi-supervised}.
By doing so, we show that the embedding of other assertions is also improved. 

\subsection{Putting it Together}

\begin{figure*}[!t]
  \centering
  \includegraphics[width=0.8\textwidth]{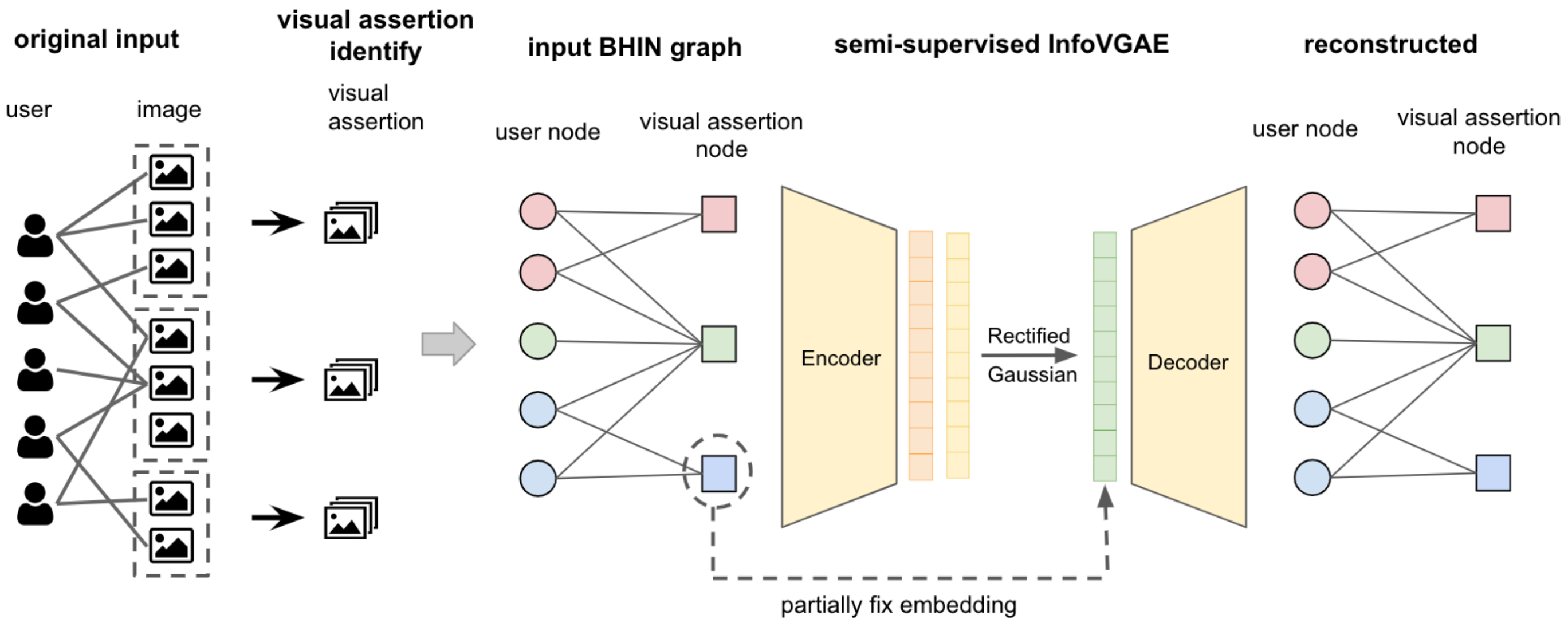}
  \caption{This is a figure showing the overall pipeline of our image ideology classification algorithm. Starting from the original user-image pair input, we firstly identify the visual assertion and convert the input data to a BHIN graph. Then we feed the BHIN graph to the semi-supervised InfoVGAE model and use the embedding result for classification.}
  \label{fig::overall_structure}
  \vspace{-5pt}
\end{figure*}

Figure~\ref{fig::overall_structure} summarizes the methodology described in this paper. Specifically, we first identify visual assertions based on the similarity among images as discussed in Section~\ref{sec:visual_model}. We then model user-assertion interactions by a bipartite graph, as described in Section~\ref{sec:graph}.  Finally, we feed the graph to the InfoVGAE for embedding, as described in Section~\ref{sec:embed} and perform ideological separation based on the embedding result. 

\section{Evaluation}
\label{sec:evaluation}
In this section, we evaluate the performance of our proposed image ideology classification algorithm. Below, we describe the data set used for evaluation, the compared baselines, and the key performance results, respectively.


\subsection{The Dataset}
While the goal of this paper is to classify {\em images\/}, for ease of data collection and ground truth estimation, we exploited Twitter as a means to download media objects. No tweet text was used by the classification algorithm, however. More specifically, we collected media objects (images) posted on Twitter about the Russia-Ukraine war from 2022-05-01 to 2022-11-02 using a keyword API prompted separately once with pro-Kremlin then once with pro-Ukraine keywords. 
For the ``pro-Ukraine" side, keywords were chosen that characterize the war as an act of aggression. For the ``pro-Kremlin" side, we used keywords associated with Kremlin-sourced messaging, including slogans (e.g., \#IstandwithPutin), allegations of Nazi influence in Ukraine, accusations that the West violated the Minsk Accords (often presented by the Russian side as a justification for the war), personal attacks on the Ukrainian president, representations of the war as self-defense against NATO, and claims of superiority of the BRICS bloc (in contrast to the G7 bloc) as a wider and more diverse representation of the World's population.  
All media content was auto-labeled as pro-Kremlin or pro-Ukraine, accordingly. After grouping similar images into visual assertions, a total of $21008$ users and $21713$ visual assertions were identified. To focus the results on ``important" content only, we filtered the resulting graph to retain only those nodes that have more than 10 edges. After filtering, $713$ users and $3097$ assertions were included in the final evaluation.

\subsection{Baselines}
In comparing the InfoVGAE-based ideological separation to the state of the art, the following baselines were used for the unsupervised and the semi-supervised scenarios, respectively. 

\subsubsection{Unsupervised Models:}
We compare our unsupervised InfoVGAE model to two baseline models under the unsupervised setting:

\begin{itemize}
    \item {\bf Non-Negative Matrix Factorization (NMF)\/}~\cite{al2017unveiling}:  This unsupervised method detects the polarization in social network by factorizing the user-assertion matrix. Unlike VGAE, where the encoder is nonlinear and the decoder is linear, the encoder and decoder in NMF are both linear.
    \item {\bf Belief Structured Matrix Factorization (BSMF)\/}~\cite{yang2020disentangling}:  This method enhances the NMF algorithm and can work well in the situations when community beliefs are partially overlapped.
\end{itemize}

\subsubsection{Semi-supervised Models:}
We compare our semi-supervised InfoVGAE model with {\em semi-supervised TIMME\/}~\cite{yang2020disentangling}, a semi-supervised multi-task and multi-relational embedding model. TIMME models the social networks as a heterogeneous graph first and then trains link prediction tasks and a latent representation task. The original TIMME model works on getting the latent representations of users, while in our task, we apply the same methodology that TIMME applies to the visual assertion network graph, and get the latent embedding for visual assertions.

\subsection{Key Performance Results}
\begin{figure*}[!ht]
     \centering
    \begin{subfigure}[b]{0.3\textwidth}
         \centering
         \includegraphics[width=\textwidth]{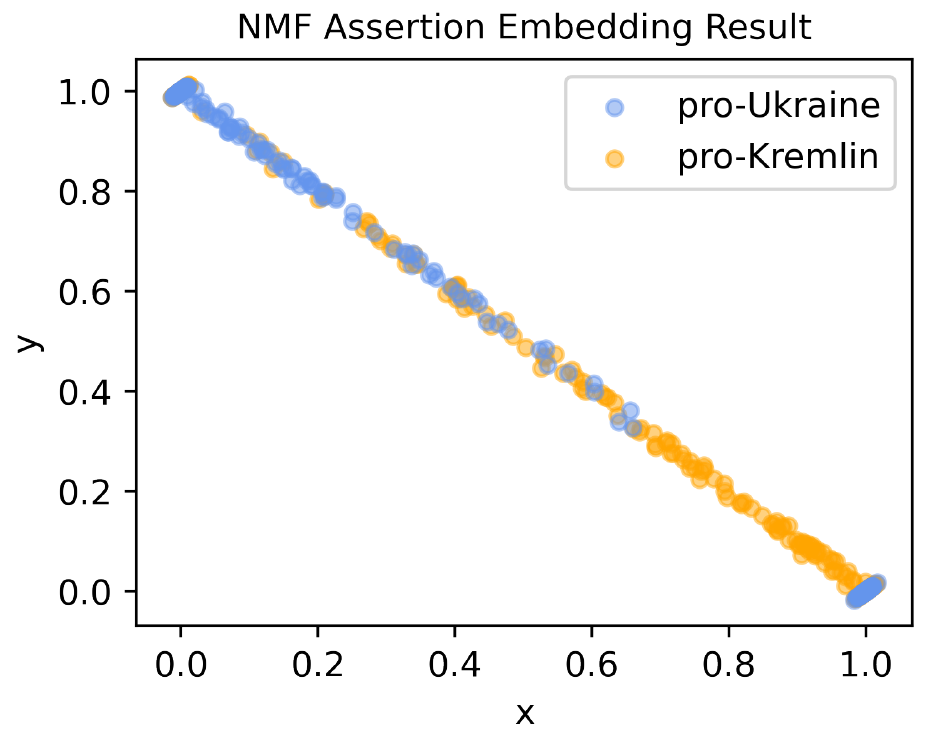}
         \caption{NMF}
         \label{fig::NMF_emb}
     \end{subfigure}
    \begin{subfigure}[b]{0.3\textwidth}
         \centering
         \includegraphics[width=\textwidth]{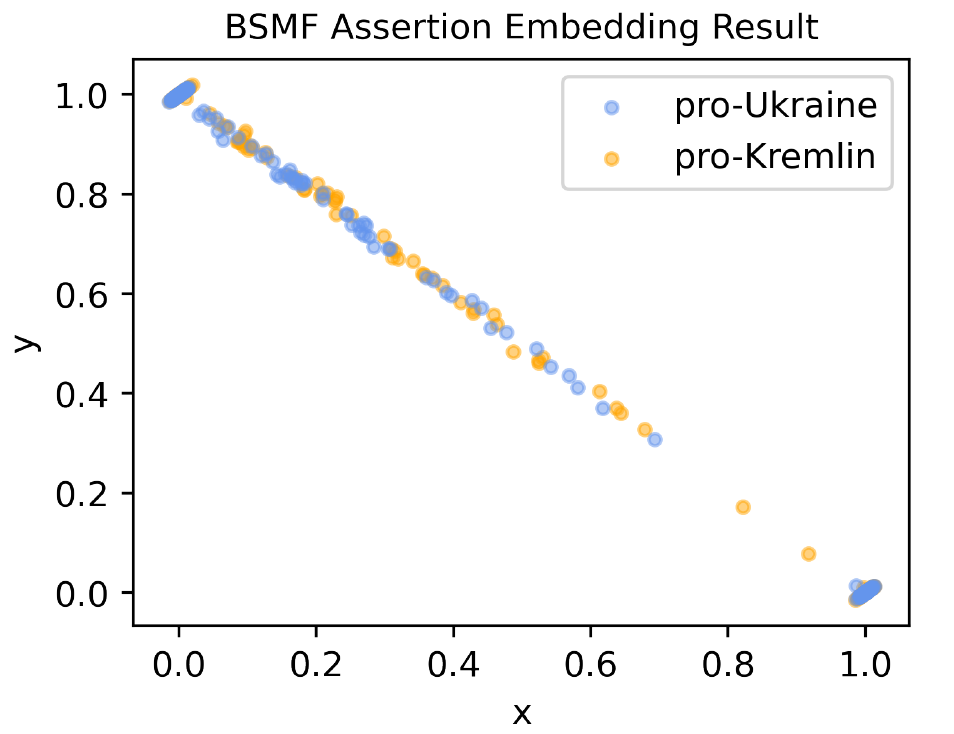}
         \caption{BSMF}
         \label{fig::BSMF_emb}
     \end{subfigure}     
     \begin{subfigure}[b]{0.3\textwidth}
         \centering
         \includegraphics[width=\textwidth]{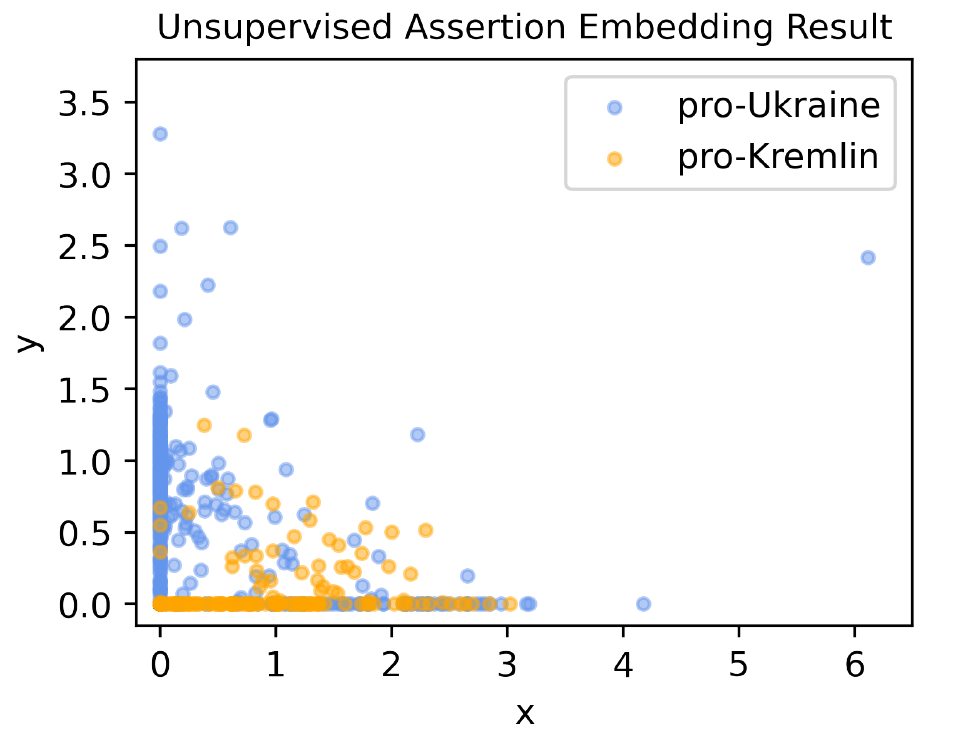}
         \caption{unsupervised InfoVGAE}
         \label{fig::unsupervised_emb}
     \end{subfigure}\\
        \begin{subfigure}[b]{0.3\textwidth}
         \centering
         \includegraphics[width=\textwidth]{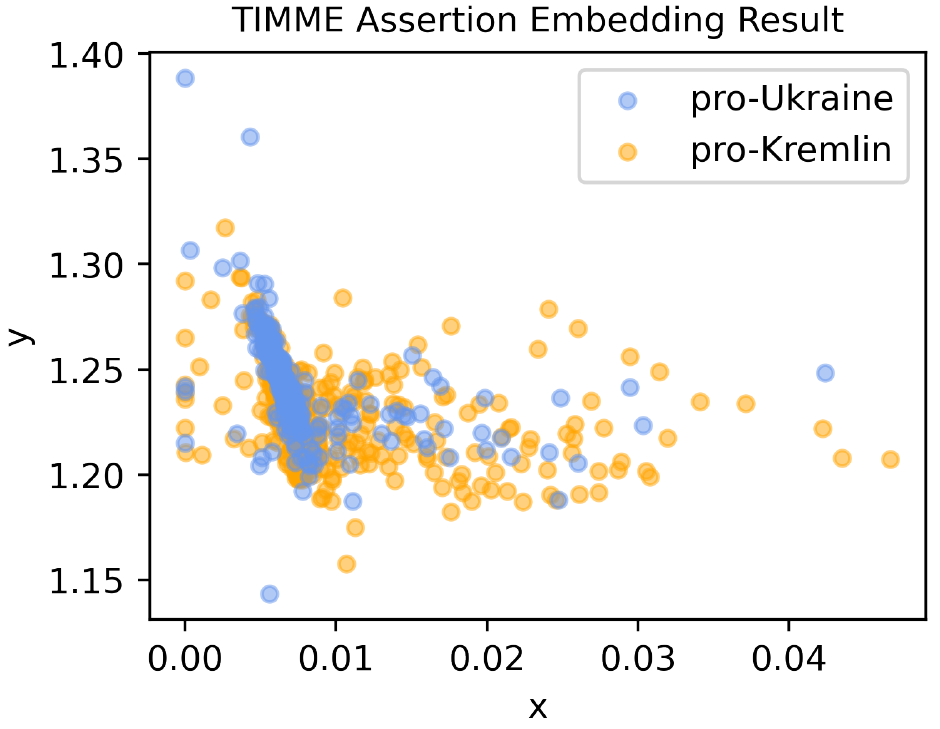}
         \caption{semi-supervised TIMME}
         \label{fig::TIMME_emb}
     \end{subfigure}
     \begin{subfigure}[b]{0.3\textwidth}
         \centering
         \includegraphics[width=\textwidth]{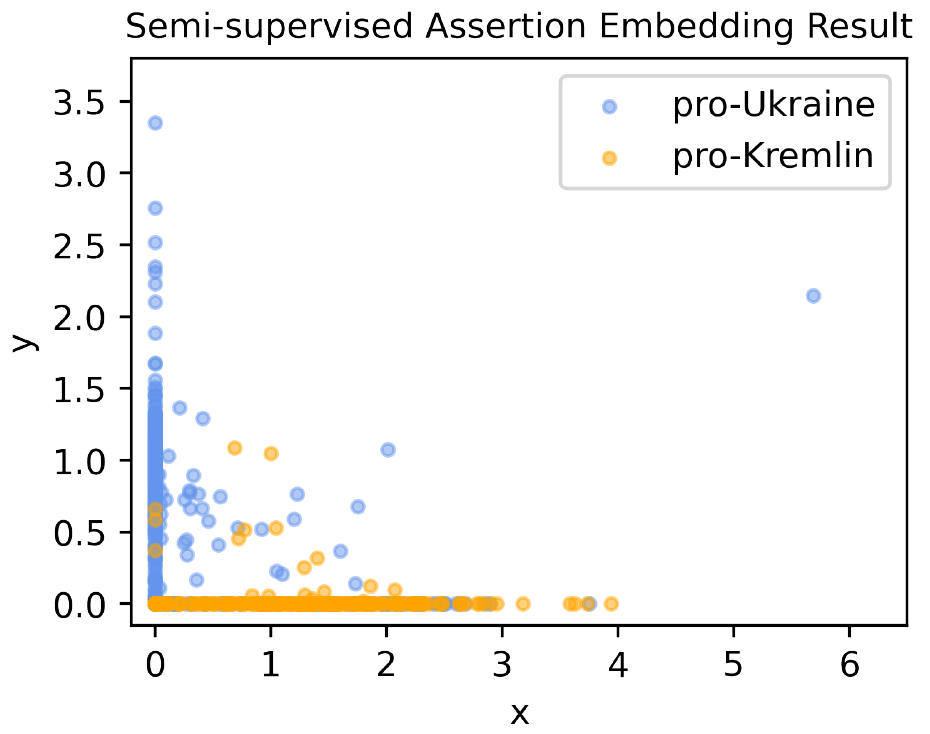}
         \caption{semi-supervised InfoVGAE}
         \label{fig::semi-supervised_emb}
     \end{subfigure}
        \caption{Comparison of assertion embedding based on unsupervised model and semi-supervised model (only the unlabeled data are visualized).}
        \label{fig::embedding_result}
\end{figure*}

Figures \ref{fig::NMF_emb}, \ref{fig::BSMF_emb}, and \ref{fig::unsupervised_emb} (embeddings on the first row of Figure~\ref{fig::embedding_result}) show the embedding results of the unsupervised approaches. The color-coding is based on ground-truth labels. Observe that the unsupervised InfoVGAE algorithm we proposed not only separates the visual assertions by ideology into largely non-overlapping clusters in the latent space but also aligns them with different axes; the $x$-axis represents a pro-Kremlin leaning, whereas the $y$-axis represents a pro-Ukraine leaning. The unsupervised algorithm does not label the axes itself, of course, but the alignment increases axis interpretability. 

Figure \ref{fig::TIMME_emb} and Figure \ref{fig::semi-supervised_emb} shows the embedding result of semi-supervised models. In both cases, 50 pro-Ukraine and 50 pro-Kremlin assertions were manually labeled (of 3097). The figures show the embedding of {\em unlabeled\/} assertions only. As can be seen from Figure~\ref{fig::TIMME_emb}, TIMME can not separate the visual assertions by ideology well when only given such a small amount of labeled data, while the semi-supervised InfoVGAE can not only separate visual assertions into different ideology groups but also ensure that the embedding of different groups aligns well with the respective axes. Note that, in this case, the user knows a priori that the $x$-axis is pro-Kremlin and the $y$-axis is pro-Ukraine. Points off the axes may be closer to neutral content, in that it is propagated to different degrees by both sides. 

From the comparison of Figure \ref{fig::unsupervised_emb} and Figure \ref{fig::semi-supervised_emb}, we can see two improvements in the embedding quality in the semi-supervised case: 1) The pro-Kremlin and pro-Ukraine visual assertions are more clearly separated. 2) The visual assertion embedding aligns better with the respective axes. 

Table~\ref{table::performance} summarizes the quantitative results, showing precision, recall, F1-score, and purity for clusters produced by the aforementioned models. The table confirms that the unsupervised InfoVGAE model we proposed achieves the best performance over other baseline models in each class (best unsupervised and best semi-supervised) on all metrics. In fact, our unsupervised algorithm also beats the semi-supervised TIMME.  
Besides, after labeling a small amount of data, the semi-supervised InfoVGAE model has a clear improvement over the unsupervised InfoVGAE model, increasing the F1 score by \textbf{5.59\%} and improving the cluster purity by \textbf{9.60\%}.

\begin{table*}[ht]
\caption{Statistical evaluation of assertion embedding performance. The evaluation compares InfoVGAE with unsupervised baseline models including NMF and BSMF, and semi-supervised baseline models including TIMME. We also compare the performance of the unsupervised and semi-supervised version of InfoVGAE.}\label{table::performance}
\centering
\small
\begin{tabular}{lccccc}
\hline
  Unsupervised Models & Precision & Recall & F1 & Purity\\
  \hline
  NMF\cite{al2017unveiling} & 0.6663 & 0.6192 & 0.6311 & 0.6192\\
  BSMF\cite{yang2020disentangling} & 0.5720 & 0.5583 & 0.5644 & 0.6377\\
  \textbf{unsupervised InfoVGAE (Ours)} & \textbf{0.7171} & \textbf{0.8735} & \textbf{0.7876} & \textbf{0.6811}\\
  \hline
    Semi-supervised Models & Precision & Recall & F1 & Purity\\
\hline
semi-supervised TIMME & 0.6147 & 0.5449 & 0.5594 & 0.5449\\
  \textbf{semi-supervised InfoVGAE (Ours)} & \textbf{0.8039} & \textbf{0.8872} & \textbf{0.8435} & \textbf{0.7771}\\
 \hline
\end{tabular}
\vspace{-5pt}
\end{table*}

Finally, we show in Table \ref{table::case_study} examples of the most popular visual assertions deemed by InfoVGAE as pro-Kremlin and those deemed as pro-Ukraine.
The table also includes brief explanations of each image. We do not claim that the beliefs expressed in these images are necessarily espoused by all members of the group in question, but simply observe that these beliefs are expressed in messaging of the corresponding side (such as messaging from the Kremlin versus messaging from the UK Ministry of Defence).  

The table illustrates the advantages of the approach used in this paper. As can be seen, the images are quite diverse; while some have overt text that reveals the stance, others are harder to interpret without proper context. By observing propagation patterns, we circumvent having to interpret the content and thus avoid reliance on complex context understanding. 

\begin{table*}[htb]
\caption{A case study of classified visual assertions into pro-Kremlin and pro-Ukraine groups.}\label{table::case_study}
\centering
\vspace{-5pt}

\begin{tabular}{|p{0.15\textwidth}|p{0.30\textwidth}|p{0.15\textwidth}|p{0.3\textwidth}|}\hline
   \multicolumn{2}{|c|}{Pro-Kremlin} & 
   \multicolumn{2}{|c|}{Pro-Ukraine} \\\hline
      Visual Assertion & Explanation & Visual Assertion & Explanation\\\hline
     \includegraphics[align=t,width=1in]{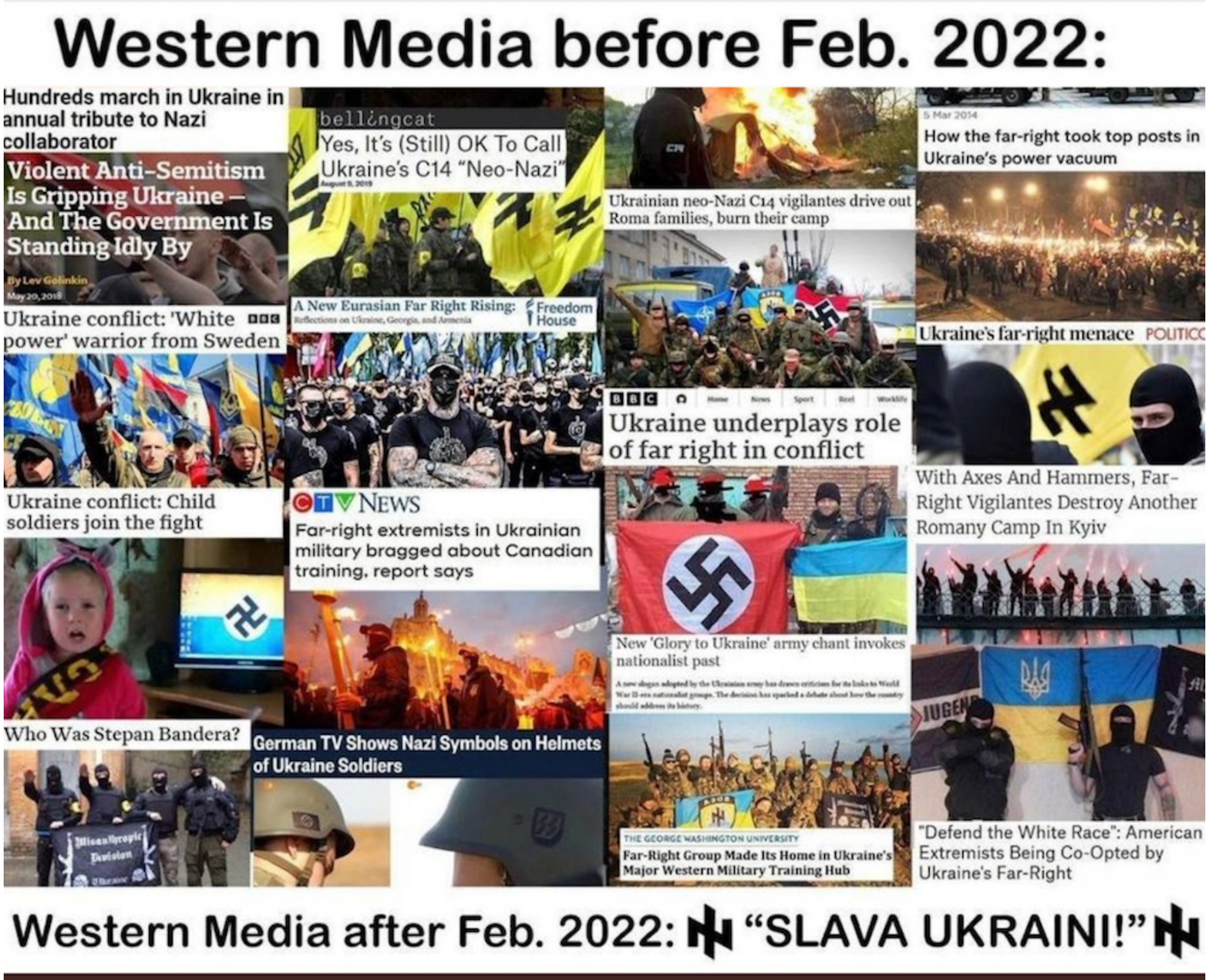}  &  An image in support of pro-Kremlin messaging that associates Ukraine with Nazi influence (depicted in the image by Nazi symbols) and satirizes Western support for Ukraine. & \includegraphics[align=t,width=1in]{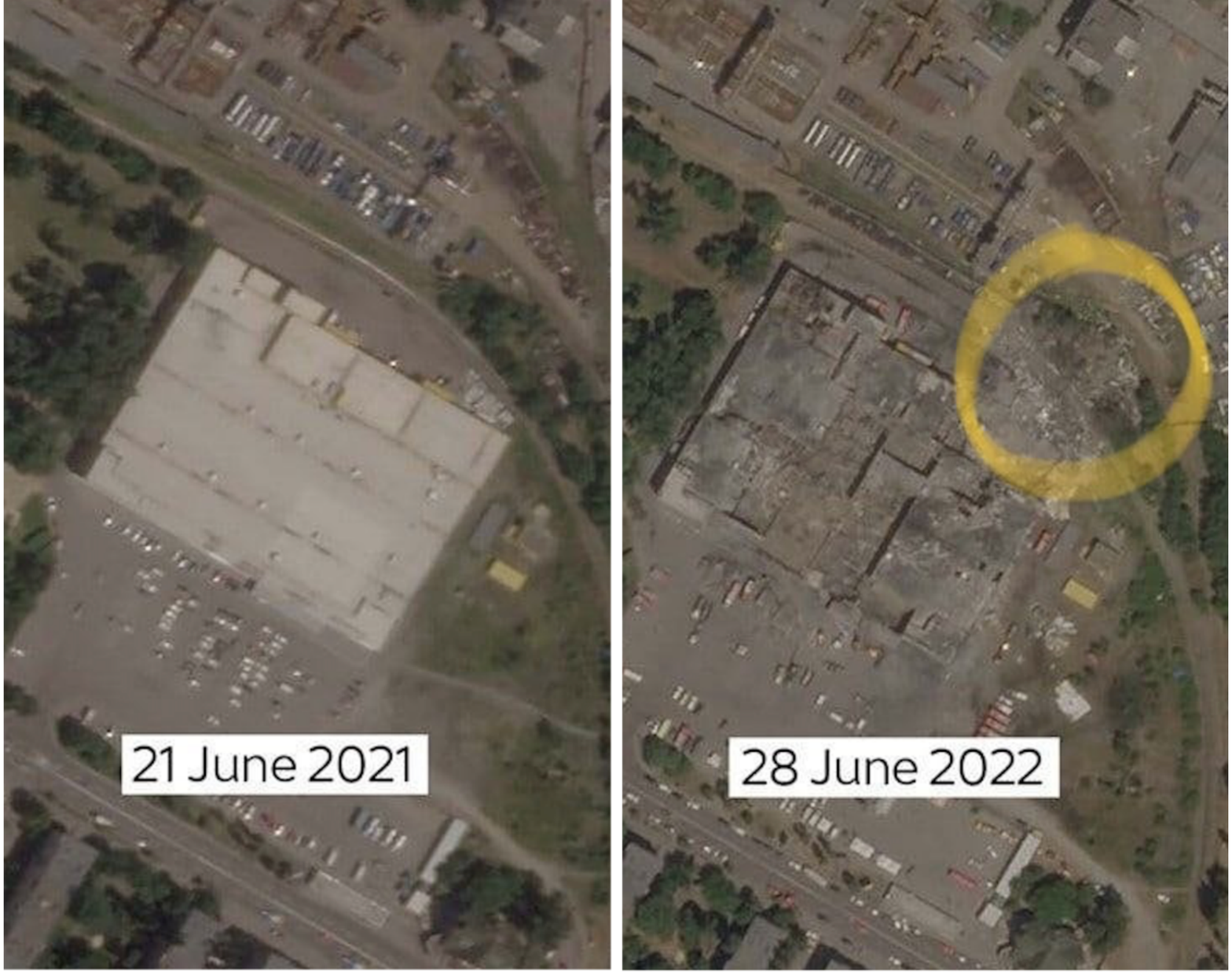}  & An image that compares  aerial photographs from June 2021 and June 2022, showing the devastating effects of Russian attacks on Ukrainian targets.\\
     \hline
     \includegraphics[align=t,width=1in]{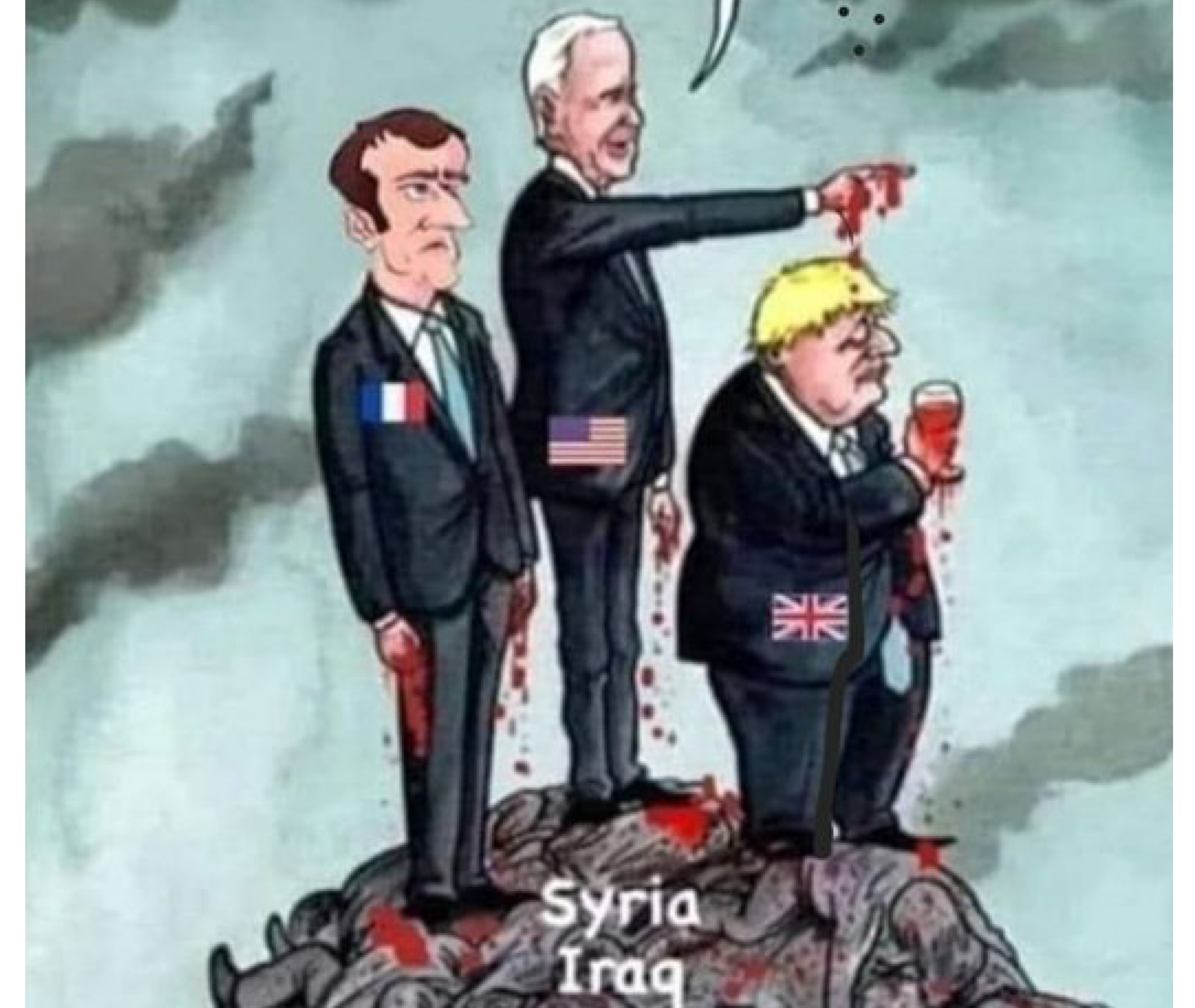}& An image with a derogatory stance towards Western leaders (France, the U.S., and the U.K.) accusing them of promoting local interests at the expense of other nations. & \includegraphics[align=t,width=1in]{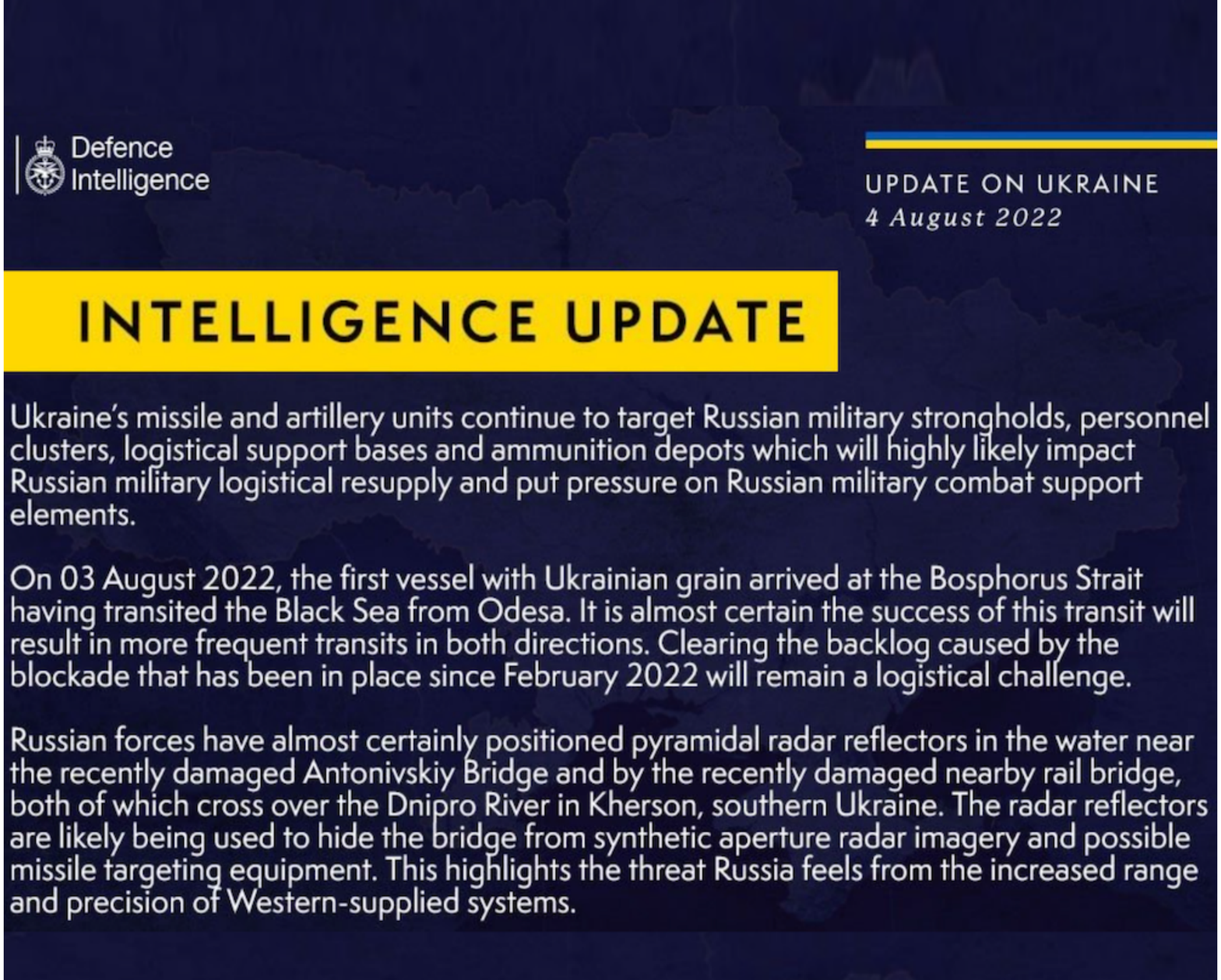}  & An image of an update by a UK intelligence agency with a pro-Ukraine stance.\\
     \hline
     \includegraphics[align=t,width=1in]{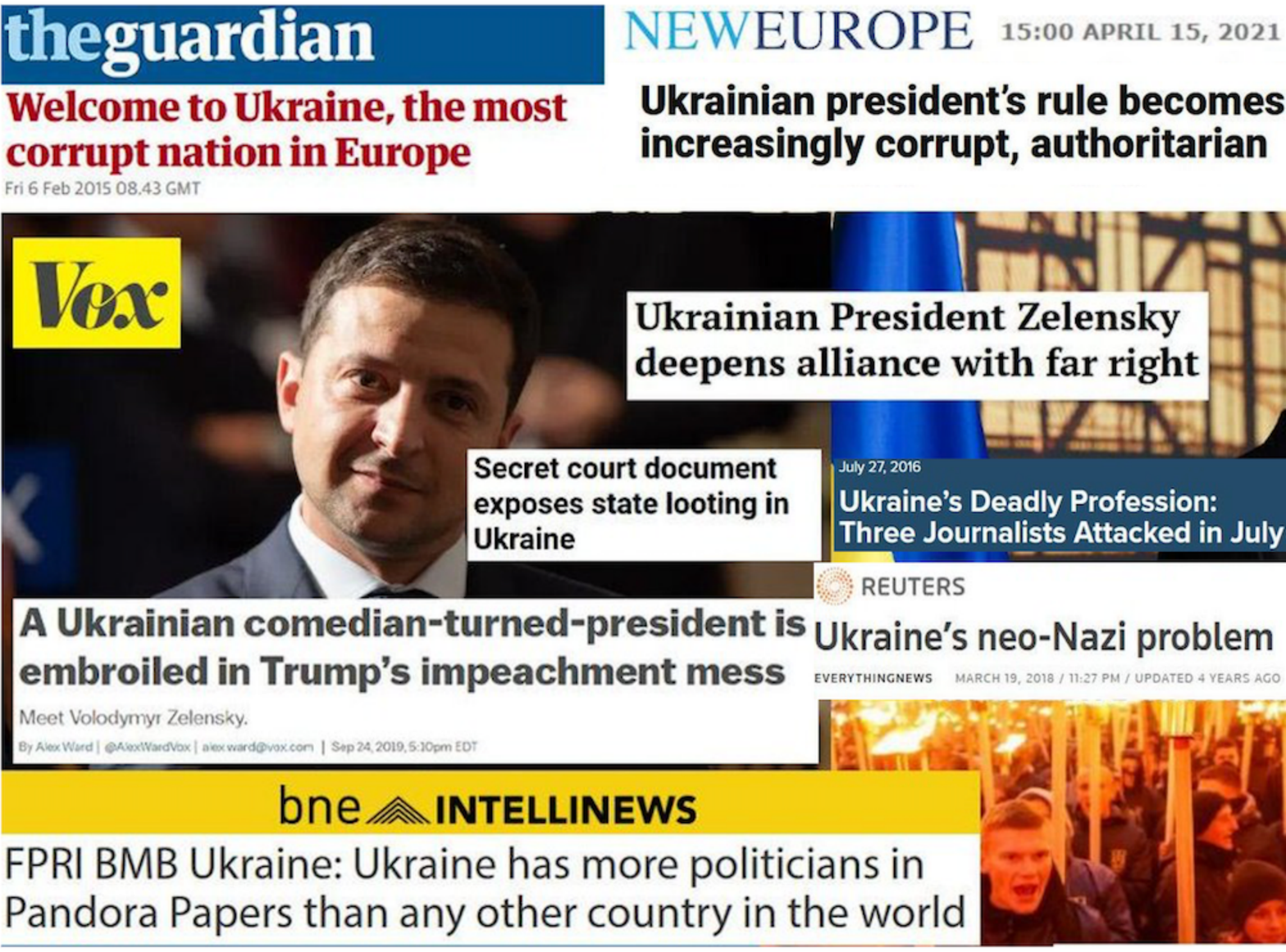}& A pro-Kremlin image whose stance is clearly evidenced by the anti-Ukraine writings on the poster. & \includegraphics[align=t,width=1in]{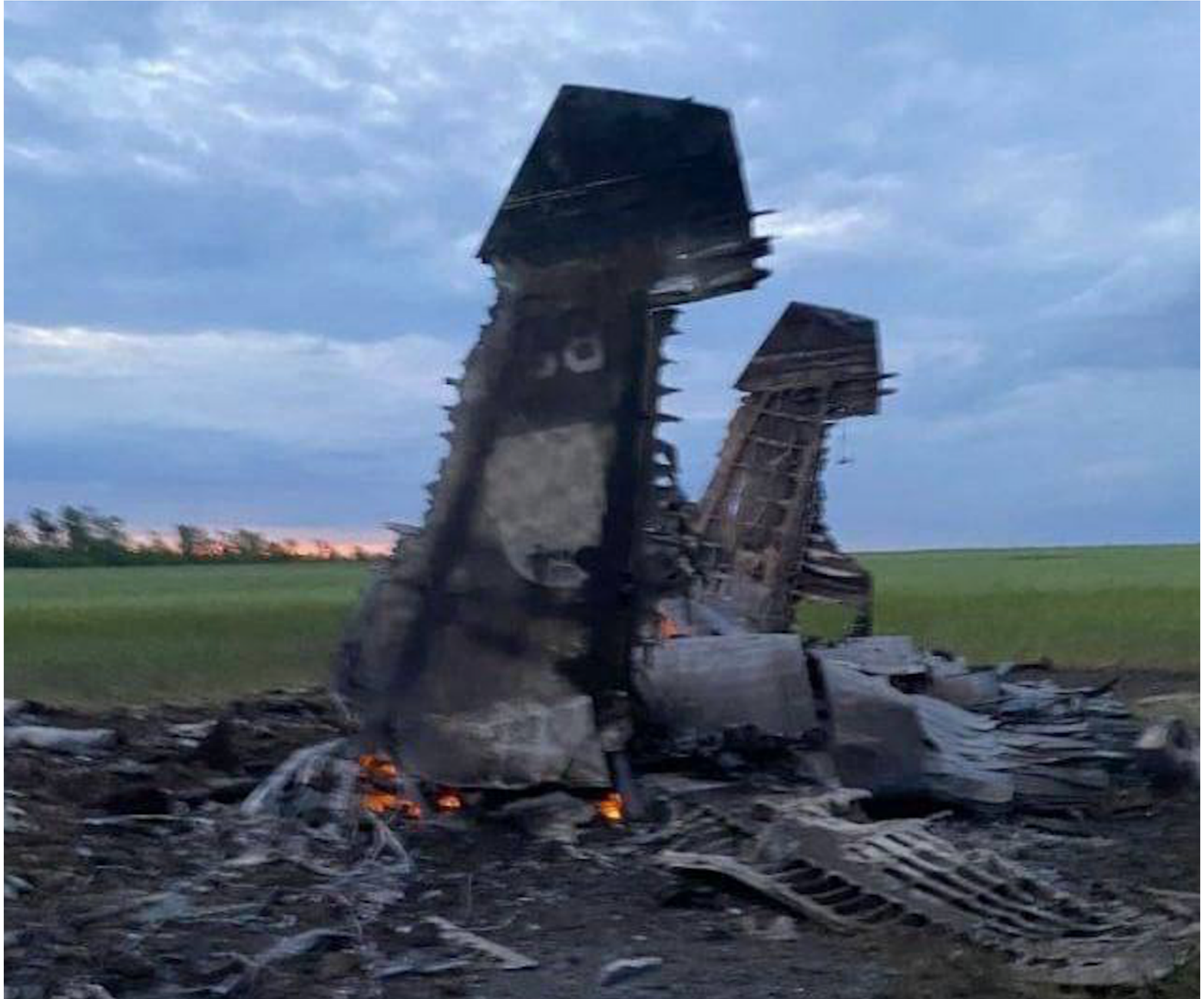}  & An image of the damage that Russian attacks have brought about in Ukraine. \\
     \hline
     \includegraphics[align=t,width=1in]{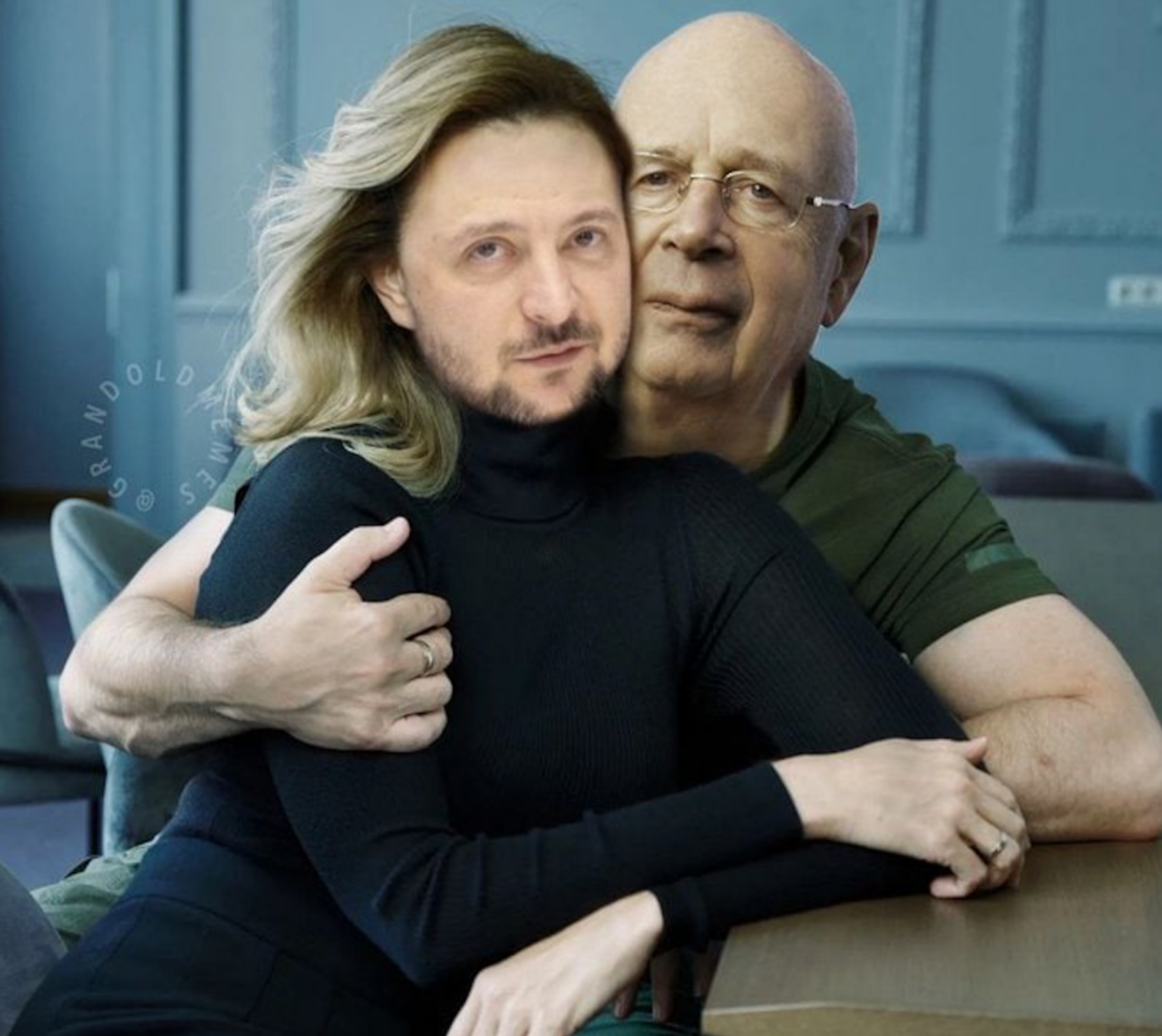}& A fake image that mocks the president of Ukraine, implying an anti-Ukraine stance. The original (from Vogue) was shown in Figure~\ref{fig:near-duplicate}. & \includegraphics[align=t,width=1in]{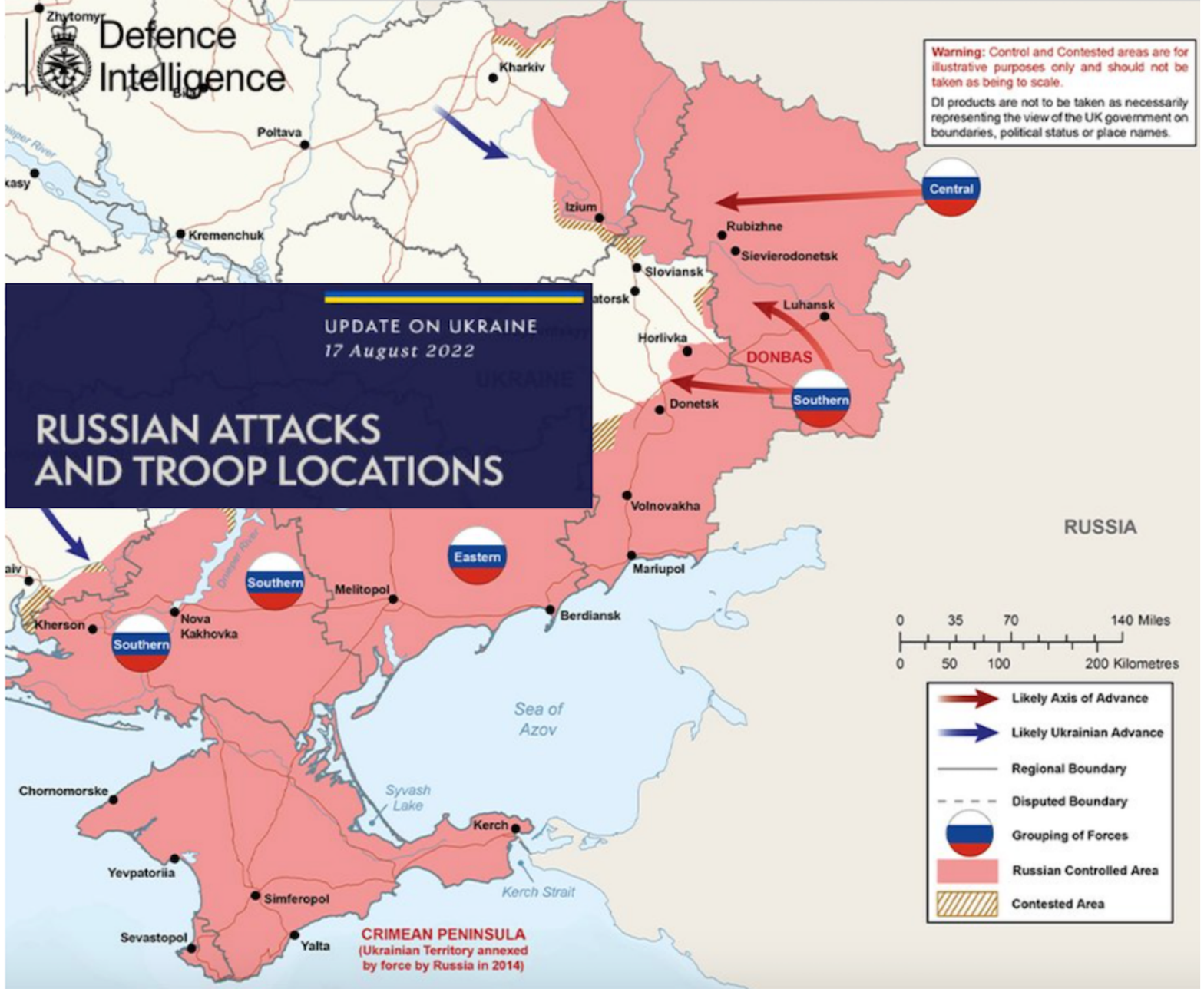} &  A map of Russian attacks and troop locations by Defence Intelligence, a UK Department of Defence agency supporting Ukraine.\\
     \hline
     \includegraphics[align=t,width=1in]{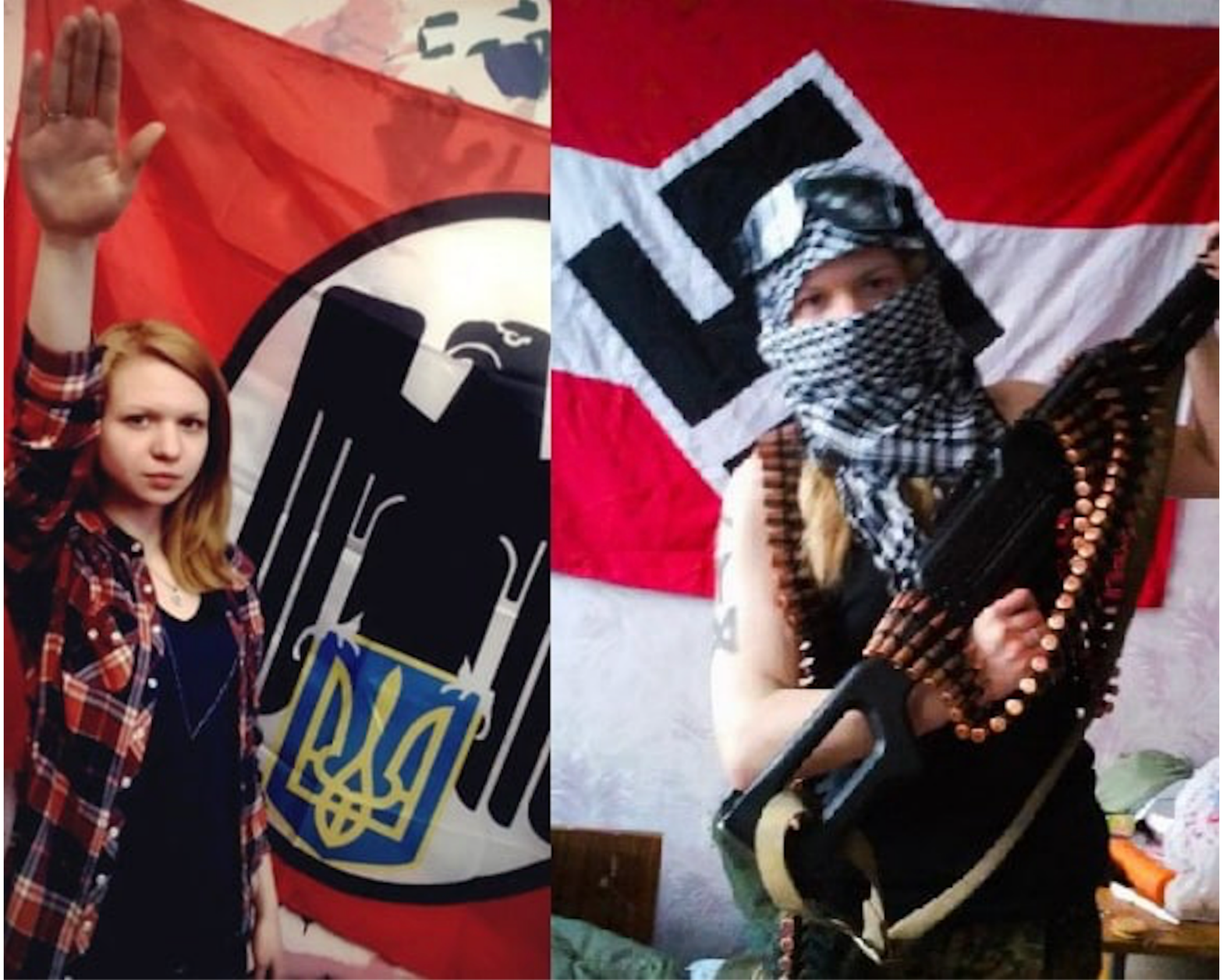}& Another image depicting an alleged neo-Nazi influence in Ukraine, a hallmark of Kremlin messaging, in an attempt to justify military activities. & \includegraphics[align=t,width=1in]{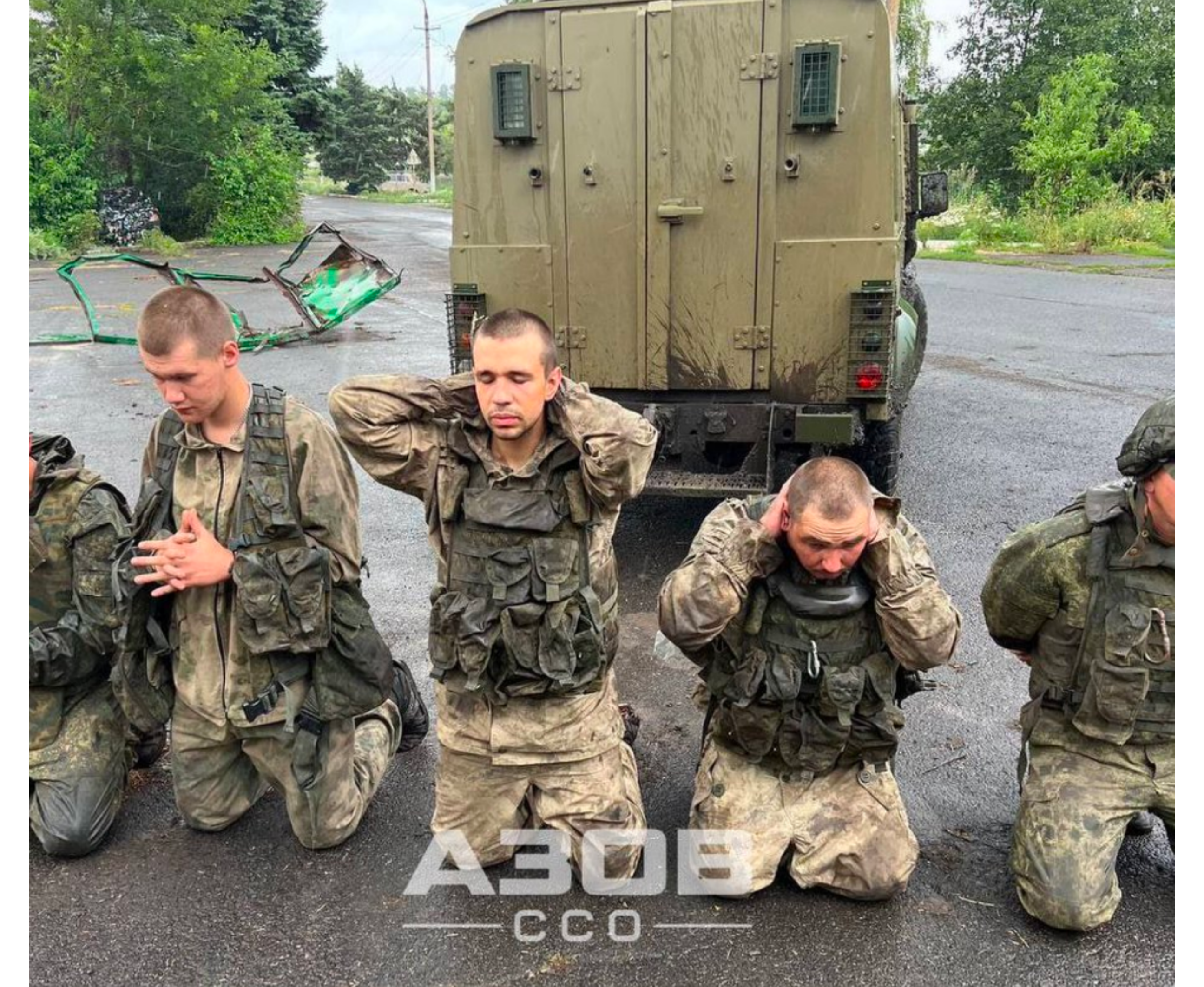}   & This image presumably shows Azov soldiers in Russian custody. It is possibly intended to convey mistreatment, although the stance of this image is unclear.\\
     \hline
     
\end{tabular}

\vspace{-5pt}
\end{table*}

\section{Discussion}
\label{sec:discussion}
The paper presented a novel unsupervised approach for classifying images by ideological alignment. While the approach shows promising results, it has several limitations that need to be highlighted. Some are logistical restrictions of the current implementation. Others are research challenges that inspire future work. Below, we discuss the key limitations.

\medskip
\noindent
{\bf Applicability restrictions:\/} The approach works best in situations involving social polarization and echo-chambers, where different groups disagree enough to interact with noticeably distinct subsets of content. There are other reasons (besides polarization) why different groups might interact with different content. For example, a group interested in cats and a group interested in race cars will access distinct content quite independently of their ideological leaning. For the solution described in this paper to work, the manner in which the original data set is curated thus becomes quite consequential. Ideally, the data set would be collected using keywords on a specific polarizing issue. Given a divisive umbrella topic, differences in content propagation patterns will likely be attributed to ideological disagreement. In the context of political debates, such polarization is (unfortunately) observed increasingly frequently. Hence, the limitation (arguably) does not pose a significant loss of applicability when it comes to classifying political imagery. 

\medskip
\noindent
{\bf Visibility requirements:\/} The approach depends on having adequate visibility into who posts what on the social medium. Social network access APIs change frequently. For example, Twitter recently restricted its access API, making it significantly more expensive to collect data at scale. Other platforms, such as Reddit and Mastodon (a new platform that has recently seen a surge of popularity in the wake of Twitter challenges) are more open. A discussion of the exact means needed for collecting the social media data is outside the scope of this work, as such a discussion is not specific to this paper. Rather, it is common to many other research directions that rely on collecting data for analysis from social~media.

\medskip
\noindent
{\bf Neutral content prevalence:\/} An important challenge to this approach is the handling of neutral content. In principle, even in a polarized debate, some imagery may be neutral and, as such, propagated by both camps. While the approach is fairly robust to having a certain amount of neutral content (InfoVGAE simply embeds it closer to the origin or near the diagonal that splits the space between two axes), some investigation is needed into the tipping point (in terms of the volume of such neutral content), after which the embedding fails to adequately separate different ideological leanings. 

\medskip
\noindent
{\bf Semi-supervised and multimodal extensions:\/} As alluded to in the paper, in many cases, while the ideological alignment of some images is hard to interpret automatically, other images are more direct. For example, a meme that explicitly labels Putin as a war criminal is easy to associate with anti-Russia sentiments using content analysis. It is therefore interesting to investigate optimal labeling strategies that maximize the efficacy of the semi-supervised solution. Such an approach is a topic for future work. Other extensions could leverage multimodal information, such as images and their captions, to improve classification accuracy.

The work is a step towards understanding the role and use of images in online political discourse, from misinformation and disinformation to political advertising and mobilization. A key to such understanding is scalable analysis. An initial step is to separate messaging of different sides of political discourse, at scale.
Below, we review some related work and outline further opportunities for extension and synergies with alternative approaches.

\section{Related Work}
\label{sec:related}
The key idea of exploiting content propagation patterns on social media for purposes of understanding the content itself was proposed in prior work under the name of {\em social sensing\/}~\cite{wang2019age}. 
Early examples included fact-finding~\cite{wang2012truth,wang2014using,wang2013recursive,shao2020mis}, event detection~\cite{wang2017storyline,gao2018dancinglines,li2019senti2pop,atefeh2015survey,DyDiff-VAE}, polarization detection~\cite{al2017unveiling,yang2020hierarchical}, and belief representation learning~\cite{li2022unsupervised}. However, previous incarnations of this idea were generally applied to the understanding of users and texts. The efficacy of the approach to classifying memes has not been systematically explored. 
 
The work described in this paper is an instance of social polarization and stance detection tasks that focus on the automatic separation of users and their posts on social media according to their polarity~\cite{tucker2018social,aldayel2021stance}. Most existing work exploits the interaction data as well as the user profile data to separate users. For example, some work applies matrix factorization on the feature matrix to successfully separate the users and their assertions \cite{al2017unveiling,yang2020hierarchical,CMH}. The authors in~\cite{darwish2020unsupervised} also explore applying the dimensional deduction methods and clustering algorithms to infer the polarity of users. In~\cite{li2022unsupervised,xiao2020timme}, the authors model the social interaction data as a graph and apply graph representation learning to understand the polarization in the beliefs of users and their posts. However, this work has not been applied to images. 

%
%

In general, computer vision techniques can be applied to solve image classification tasks. 
Convolutional neural networks (CNNs)~\cite{simonyan2014very,he2016deep,krizhevsky2017imagenet} and more modern architectures such as vision transformers have been widely successful in a broad variety of classification tasks but would require topic-specific annotations for fine-tuning for the problem discussed in this paper. 
While existing work applied CNNs to image sentiment detection~\cite{you2015robust}, object identification~\cite{zhao2019object}, and segmentation~\cite{minaee2021image}, it does not directly allow ideological classification of images due to the need for understanding additional political context. In~\cite{xi2020understanding}, the author combines the party affiliation classification of politicians' photos and facial sentiment detection techniques to classify the ideology of images. However, the performance of this model is limited because it is very difficult to generalize to photos of unseen politicians as well as other non-politician people, and therefore suffers from the over-fitting problem. Recent  large vision-language models such as CLIP \cite{radford2021CLIP} are one avenue to inject more implicit contextual knowledge into (zero-shot) image classification, but ultimately always hinge on their ability to instill meaning into the content of images. In contrast, the proposed approach derives meaning solely from the interaction data between social entities and the images shared among them. 

Graph representation techniques are proven to be effective in detecting ideology. Most existing graph-based techniques construct a graph using the historical interaction data on social networks and learn representations from it~\cite{wang2018ace,DyDiff-VAE,wang2022,wang2022learning,wang2023mpkd}. InfoVGAE~\cite{li2022unsupervised} constructs a heterogeneous bipartite graph and learns the unsupervised belief representations of users and tweets with the property of ideological separation. In~\cite{gu2017ideology}, the authors model the ideology of users with a message propagation formula on heterogeneous types of social links in an unsupervised manner. In~\cite{akoglu2014quantifying}, the author constructs a signed bipartite network and formulates the polarity analysis as an unsupervised link classification task. However, most existing graph-based models only consider the users and posts as nodes. They have not been tested on multi-media memes. In this paper, we propose to classify the multi-media memes with a graph containing users and multi-media memes. We also explore enhancing the performance of classification with some minimal supervision (or multi-media meme annotations).

\section{Conclusions}
\label{sec:conclusions}
The work is a step towards improving the understanding of visual memes used in political discourse. The approach can be thought of as implicitly relying on a form of crowd-sourcing, where individuals use their cognitive skills to interpret messages then act accordingly (e.g., like, up-vote, forward, or ignore the meme). Their collective actions result in a propagation pattern unique to the meme's content. This propagation pattern is then used to help distinguish different ideological leaning in memes. Evaluation has shown that the general idea of leveraging propagation patterns of memes in the population is a promising approach that deserves more systematic investigation. The discussion section identified limitations and further avenues for future work. The authors will pursue such opportunities in future publications.

\section*{Acknowledgments}
Research reported in this paper was sponsored in part by DARPA awards HR001121C0165 and HR00112290105, the DoD Basic Research Office award HQ00342110002, and the Army Research Laboratory under Cooperative Agreement W911NF17-20196.

\balance

\bibliographystyle{aaai}
\bibliography{paper}

\end{document}